\documentclass[iicol]{sn-jnl}

\usepackage{graphicx}%
\usepackage{multirow}%
\usepackage{amsmath,amssymb,amsfonts}%
\usepackage{amsthm}%
\usepackage{mathrsfs}%
\usepackage[title]{appendix}%
\usepackage{xcolor}%
\usepackage{textcomp}%
\usepackage{manyfoot}%
\usepackage{booktabs}%
\usepackage{algorithm}%
\usepackage{algorithmicx}%
\usepackage{algpseudocode}%
\usepackage{listings}%
\usepackage{todonotes}
\usepackage{comment}

\usepackage{color}
\definecolor{lightgray}{gray}{0.85}

\newcommand\greybox[1]{%
  \vskip\baselineskip%
  \par\noindent\colorbox{lightgray}{%
    \begin{minipage}{.97\columnwidth}#1\end{minipage}%
  }%
  \vskip\baselineskip%
}

\begin{document}

\title[Human Factors in Model-Driven Engineering]{Human Factors in Model-Driven Engineering: Future Research Goals and Initiatives for MDE}

\author*[1]{\fnm{Grischa} \sur{Liebel}}\email{grischal@ru.is}
\author*[2]{\fnm{Jil} \sur{Klünder}}\email{jil.kluender@inf.uni-hannover.de}
\author*[3]{\fnm{Regina} \sur{Hebig}}\email{regina.hebig@uni-rostock.de}
\author[4]{\fnm{Christopher} \sur{Lazik}}\email{lazikchr@informatik.hu-berlin.de}
\author[5]{\fnm{Inês} \sur{Nunes}}\email{ir.nunes@campus.fct.unl.pt}
\author[6]{\fnm{Isabella} \sur{Graßl}}\email{isabella.grassl@uni-passau.de}
\author[7]{\fnm{Jan-Philipp} \sur{Steghöfer}}\email{jan-philipp.steghoefer@xitaso.com}
\author[8]{\fnm{Joeri} \sur{Exelmans}}\email{joeri.exelmans@uantwerpen.be}
\author[3]{\fnm{Julian} \sur{Oertel}}\email{julian.oertel@uni-rostock.de}
\author[9]{\fnm{Kai} \sur{Marquardt}}\email{kai.marquardt@kit.edu}
\author[10]{\fnm{Katharina} \sur{Juhnke}}
\email{katharina.juhnke@zeiss.com}
\author[2]{\fnm{Kurt} \sur{Schneider}}\email{kurt.schneider@inf.uni-hannover.de}
\author[11]{\fnm{Lucas} \sur{Gren}}\email{lucas.gren@gu.se}
\author[12]{\fnm{Lucia} \sur{Happe}}\email{lucia.happe@kit.edu}
\author[2]{\fnm{Marc} \sur{Herrmann}}\email{marc.herrmann@inf.uni-hannover.de}
\author[13]{\fnm{Marvin} \sur{Wyrich}}\email{wyrich@cs.uni-saarland.de}
\author[14]{\fnm{Matthias} \sur{Tichy}}
\email{matthias.tichy@uni-ulm.de}
\author[15]{\fnm{Miguel} \sur{Goulão}}\email{mgoul@fct.unl.pt}
\author[11]{\fnm{Rebekka} \sur{Wohlrab}}\email{wohlrab@chalmers.se}
\author[16]{\fnm{Reyhaneh} \sur{Kalantari}}\email{Reyhaneh.kalantari@uottawa.ca}
\author[17]{\fnm{Robert} \sur{Heinrich}}\email{robert.heinrich@kit.edu}
\author[18]{\fnm{Sandra} \sur{Greiner}}\email{sandra.greiner@unibe.ch}
\author[19]{\fnm{Satrio Adi} \sur{Rukmono}}
\email{s.a.rukmono@tue.nl}
\author[1]{\fnm{Shalini} \sur{Chakraborty}}\email{shalini19@ru.is}
\author[20]{\fnm{Silvia} \sur{Abrahão}}\email{sabrahao@dsic.upv.es}
\author[15]{\fnm{Vasco} \sur{Amaral}}\email{vasco.amaral@fct.unl.pt}

\affil*[1]{\orgdiv{Department of Computer Science}, \orgname{Reykjavik University}, \orgaddress{\street{Menntavegur 1}, \city{Reykjavik}, \postcode{102}, \country{Iceland}}}

\affil*[2]{\orgdiv{Software Engineering Group}, \orgname{Leibniz University Hannover}, \orgaddress{\street{Welfengarten 1}, \city{Hannover}, \postcode{30167}, \country{Germany}}}

\affil*[3]{\orgdiv{Department of Software Engineering}, \orgname{University of Rostock}, \orgaddress{\street{Albert-Einstein-Straße}, \city{Rostock}, \postcode{18057}, \country{Germany}}}

\affil[4]{\orgdiv{Department of Software Engineering}, \orgname{Humboldt-Universität zu Berlin}, \orgaddress{\street{Unter den Linden 6}, \postcode{10099}, \city{Berlin}, \country{Germany}}}

\affil[5]{\orgname{Unaffiliated Researcher}, \orgaddress{\city{Frankfurt}, \country{Germany}}}

\affil[6]{\orgdiv{Department of Software Engineering II}, \orgname{University of Passau}, \orgaddress{\street{Innstraße 33}, \city{Passau}, \postcode{94032}, \country{Germany}}}

\affil[7]{\orgname{XITASO GmbH}, \orgaddress{\street{Austraße 35}, \postcode{86153}, \city{Augsburg}, \country{Germany}}}

\affil[8]{\orgdiv{Department of Computer Science}, \orgname{University of Antwerp}, \orgaddress{\street{Middelheimlaan 1}, \postcode{2020}, \city{Antwerp}, \country{Belgium}}}

\affil[9]{\orgdiv{Department of Informatics}, \orgname{Karlsruhe Institute of Technology}, \orgaddress{\street{Am Fasanengarten 5} \city{Karlsruhe}, \postcode{75131}, \country{Germany}}}

\affil[10]{\orgdiv{SMT-EMI6}, \orgname{Carl Zeiss SMT GmbH}, \orgaddress{\street{Rudolf-Eber-Straße 2}, \postcode{73447}, \city{Oberkochen}, \country{Germany}}}

\affil[11]{\orgdiv{Department of Computer Science and Engineering}, \orgname{Chalmers $|$ University of Gothenburg}, \orgaddress{
\city{Gothenburg}, \postcode{41127}, \country{Sweden}}}

\affil[12]{\orgdiv{Department of Interdisciplinary Didactics}, \orgname{Karlsruhe Institute of Technology}, \orgaddress{\street{Engesserstr. 2} \city{Karlsruhe}, \postcode{75131}, \country{Germany}}}

\affil[13]{\orgdiv{Department of Computer Science}, \orgname{Saarland University}, \orgaddress{\street{Saarland Informatics Campus}, \city{Saarbrücken}, \postcode{66123}, \country{Germany}}}

\affil[14]{\orgdiv{Institute of Software Engineering and Languages}, \orgname{Ulm University}, \orgaddress{\street{James-Franck-Ring}, \postcode{89081}, \city{Ulm}, \country{Germany}}}

\affil[15]{\orgdiv{Department of Computer Science}, \orgname{NOVA School of Science \& Technology}, \street{Campus de Caparica}, \postcode{2829-516} \city{Caparica}, \country{Portugal}}

\affil[16]{\orgdiv{Department of Engineering Design and Teaching Innovation}, \orgname{University of Ottawa}, \orgaddress{\street{800 King Edward Avenue}, \city{Ottawa}, \postcode{K1N6N5}, \country{Canada}}}

\affil[17]{\orgdiv{KASTEL - Institute of Information Security and Dependability}, \orgname{Karlsruhe Institute of Technology}, \orgaddress{\street{Am Fasanengarten 5} \city{Karlsruhe}, \postcode{75131}, \country{Germany}}}

\affil[18]{\orgdiv{Department of Computer Science}, \orgname{University of Bern}, \orgaddress{
\city{Bern}, \postcode{3012}, \country{Switzerland}}}

\affil[19]{\orgdiv{Department of Mathematics and Computer Science}, \orgname{Eindhoven University of Technology}, \orgaddress{\street{De Zaale}, \postcode{5600 MB}, \city{Eindhoven}, \country{The Netherlands}}}

\affil[20]{\orgdiv{Instituto Universitario Mixto de Tecnología Informática}, \orgname{Universitat Politècnica de València}, \street{Camino de Vera s/n}, \postcode{46022} \city{Valencia}, \country{Spain}}

 \abstract{\textbf{Purpose:} Software modelling and Model-Driven Engineering (MDE) is traditionally studied from a technical perspective. However, one of the core motivations behind the use of software models is inherently human-centred. Models aim to enable practitioners to communicate about software designs, make software understandable, or make software easier to write through domain-specific modelling languages. Several recent studies challenge the idea that these aims can always be reached and indicate that human factors play a role in the success of MDE.
However, there is an under-representation of research focusing on human factors in modelling.
 
\textbf{Methods:} During a GI-Dagstuhl seminar, topics related to human factors in modelling were discussed by 26 expert participants from research and industry.
 
 \textbf{Results:}  In breakout groups, five topics were covered in depth, namely modelling human aspects, factors of modeller experience, diversity and inclusion in MDE, collaboration and MDE, and teaching human-aware MDE.
 
 \textbf{Conclusion:} 
 We summarise our insights gained during the discussions on the five topics. We formulate research goals, questions, and propositions that support directing future initiatives towards an MDE community that is aware of and supportive of human factors and values.
 }


\maketitle

\section{Introduction}\label{sec1}
The use of software models to design, analyse, generate, and document software, often called Model-Driven Engineering (MDE), is at the core of software engineering (SE) research and practice.
Strong communities have been built around modelling technologies, as evidenced by several international conferences specialising in the topic, e.g., the CORE A-ranked Models conference\footnote{\url{https://modelsconference.org}}.

Models are aimed at enabling practitioners to communicate about software designs~\cite{jolak2020software}, at making software understandable~\cite{ho2017practices}, or at making software easier to write through domain-specific modelling languages~\cite{gotz2021claimed}.
The modelling community traditionally studies these aspects from a technical perspective, considering aspects such as the technical design of language workbenches, specifications of modelling standards, grammar generation from meta models, and model transformation languages.
However, one of the core motivations behind the use of software models is inherently human-centred, and several recent studies indicate that human factors play a role in the success of MDE~\cite{liebel2019use,hoppner2022advantages}.
Such research focusing on human factors in modelling is currently under-represented, including topics such as:
\begin{itemize}
    \item human factors that influence the way models are used;
    \item design of modelling languages and technologies that account for human factors, such as cognitive ergonomics and accessibility issues;
    \item the potential of models in supporting humans in SE, e.g., during software comprehension, for explainable AI, or usability design.
\end{itemize}

To address this gap in existing work, we gathered a group of 26 researchers related to human factors and modelling research in Schloss Dagstuhl in November 2023 to constructively discuss ways forward in research related to human factors in MDE\footnote{\url{https://www.dagstuhl.de/23473}}.
In the following, we summarise the discussions during the seminar, as well as the recommendations and suggested actions resulting from them.

\section{Overview of Discussed Topics}
\label{sec:topicOverview}
The insights presented in this paper emerge from discussions at a GI-Dagstuhl seminar on ''Human Factors in Model-Driven Engineering''\footnote{\url{https://www.dagstuhl.de/23473}}. Details about the participants of the seminar can be found in the appendix \ref{sec:seminar}. 

Using the 1-2-4-ALL microstructure\footnote{\url{https://www.liberatingstructures.com/1-1-2-4-all/}}, we elaborated goals, outcomes, and topics to be discussed over the week.
In this practice, participants are encouraged to individually brainstorm specific topics for a limited amount of time, such as the goals, outcomes, and topics for the seminar.
This step is followed by pairing up with another participant, comparing the individual results and agreeing on joint results.
The same is repeated with groups of four and, finally, with all participants.

In the first step, the participants elicited their goals and expected outcomes of the workshop with the 1-2-4-ALL microstructure.
The most frequently mentioned (and aligned) goals were (1) to formulate a research agenda respectively to build a roadmap, (2) to exchange ideas, (3) to make MDE more practical, (4) to unify projects, (5) to develop ideas for empirical research on MDE in industry, and (6) to use more models (of humans). 

The requested outcomes of the workshop were quite manifold, but all participants agreed upon establishing contacts and collaborations for specific subtopics and finding new friends. From a research perspective, the mentioned outcomes included writing joint papers, research proposals, or a book; finding approaches to integrate human factors in software engineering; to establish concepts to integrate MDE in development workflows; to identify best practices for research in industrial settings; to find strategy guidelines for teaching MDE; and to develop a new paradigm for human behaviour. 

In the next step, the group of participants identified a total of ten topics to be discussed over the week:
\begin{enumerate}
    \item Factors of Modeller Experience
    \item Collaboration and MDE
    \item Diversity and Inclusion in MDE
    \item Modelling Human Factors
    \item Teaching Human-aware MDE
    \item Research Methods
    \item AI complementing MDE
    \item MDE ecosystem maturity
    \item Usability in live modelling
    \item Sustainability and well-being
\end{enumerate}

As ten topics were considered too many to be discussed in detail over a week, and as the number of goals and outcomes was too large, a voting with three votes per participant helped prioritise the topics. This resulted in the final five topics (Topics 1 to 5 in the list above) being discussed throughout the week. The same voting also allowed us to condense the outcomes into:

\begin{itemize}
    \item New active contacts
    \item Concepts to embed MDE in development workflows
    \item Best practices for research in industrial settings
    \item A collaborative paper
\end{itemize}

The identified two main goals are (1) to elicit ideas for empirical research on MDE in industry and (2) to use more models of and for humans. 

In the following sections, we will describe the outcomes of the discussions surrounding each topic in more detail.

\section{Topic 1: Factors of Modeller Experience}
\label{sec:topic1}

In the breakout group for Topic 1, we discussed the factors that
contribute to the Modeller Experience (MX). 
The motivation was to recognise MX's key role in MDE adoption and its lack of detailed definition in scholarly work. 
Our discussion aimed to systematise MX by identifying and defining its contributing factors within MDE.

A fundamental insight is that MX varies in organisations' modelling contexts. For example, those using formal MBSE approaches, such as in the automotive sector, differ markedly in their modelling practices compared to organisations using informal modelling processes. This has led us to recognise four key modelling workflows, each reflecting distinct aspects of MX in industrial environments: 1) Low-code; 2) Everything-as-code; 3) MBSE in the Automotive Domain; and 4) Informal Modelling. We have only considered the last three of these workflows in this work.

The main objective of this work is to address the following questions: What factors contribute to MX in the different modelling workflows identified? By answering this question, we hope to better understand the needs of practitioners based on their type of workflow and, consequently, be able to provide better solutions in our research. We also believe that characterising modelling solutions using our workflows and factors will make it easier to define a target audience and delineate the research scope.

To answer this question, a focus group of ten modelling experts from academia and industry brainstormed using the `1-2-4-all' technique to identify factors more tailored towards the modeller, such as required training, maintainability, immediate benefits, and integration into the programming ecosystem. The participants then discussed how these different factors were related to each other. 

\begin{figure*}[tb]
    \centering
    \includegraphics[width=.8\linewidth]{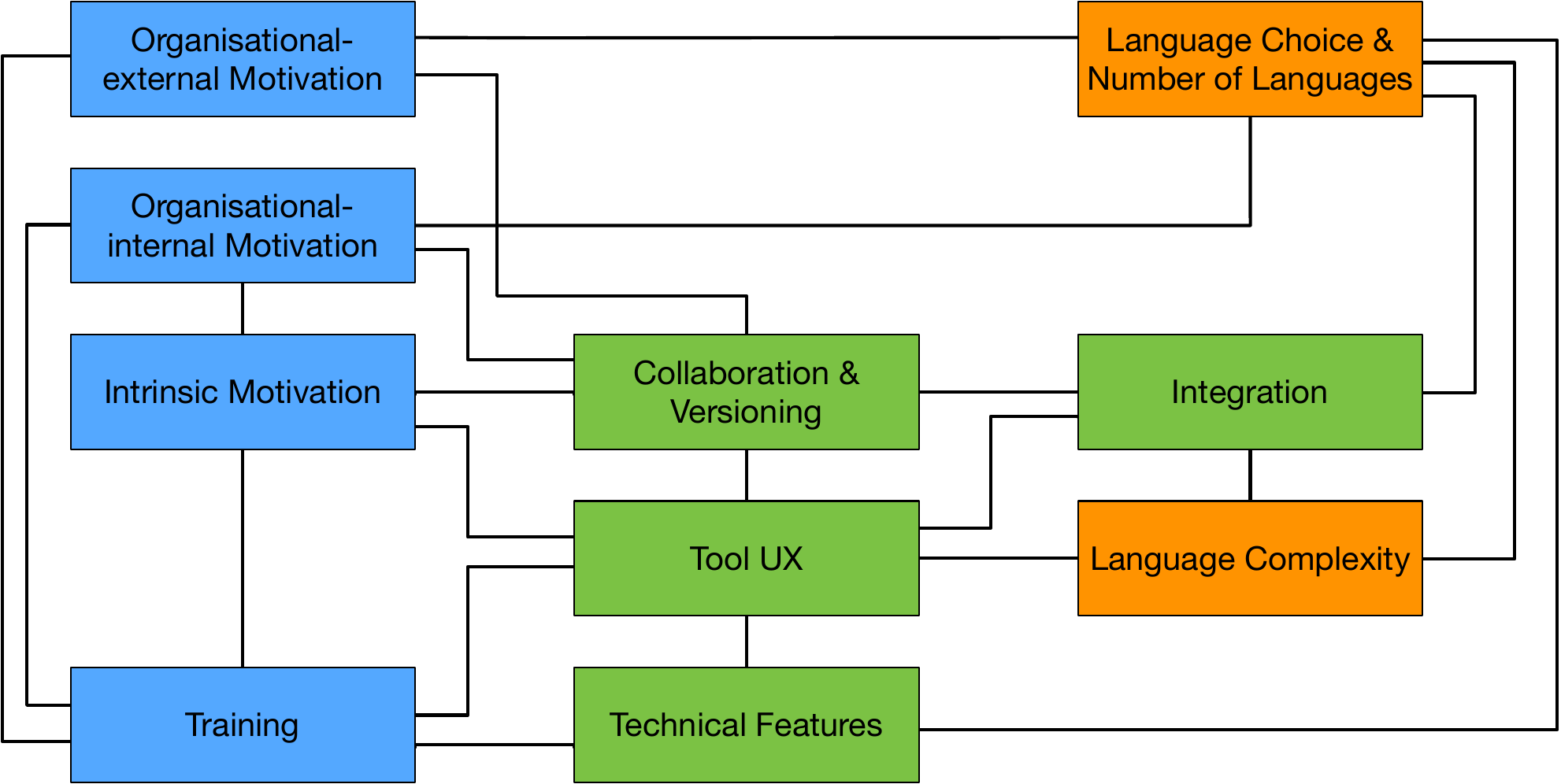}
    \caption{Factors of modeller experience and their relationships. Inherent factors are shown in orange, technical factors in green, and non-technical ones in blue.}
    \label{fig:mx-factors}
\end{figure*}

The result of the intense discussions is the set of factors shown in Figure~\ref{fig:mx-factors}. They are split into inherent factors (which are inevitable based on the characteristics of the addressed problem), technical factors and non-technical factors and are described as follows.

\paragraph{Inherent factors}

\begin{description}
    \item [\textbf{Language Complexity}:] Measures the effort to solve modelling problems using a language. Perceived complexity impacts adoption, as suggested by France et al. ~\cite{France2007Seminal}, who highlight the challenge of managing language complexity in practice.
    
    \item [\textbf{Language Choice}:] Involves selecting, extending, or defining modelling languages that provide suitable domain-specific abstractions, considering factors like vendor lock-in, which affects interoperability. The chosen language directly influences MX, as highlighted in literature reviews~\cite{kalantari2022slr}. Language choice is influenced by the system domain, the ``sphere of knowledge, influence, or activity''~\cite{evans2014domain} the system works in. A modelling language also covers several viewpoints, i.e., abstractions that specify the system restricted to certain concerns, created with specific purposes~\cite{gonzalez2013defining}.
\end{description}

\paragraph{Technical factors}

\begin{description}
    \item [\textbf{Integration}:] Refers to the capability of modelling approaches to align with existing development processes and platforms. It involves the ease of chaining multiple tools together for end-to-end functionalities and the flexibility of tools to adapt to various processes without requiring additional methods or tools~\cite{Whittle2015Taxonomy}.
    
    \item [\textbf{Tool UX}:] Concerns users' perceptions and responses that result from the use and/or anticipated use of modelling tools~\cite{iso9241-210}. It includes perceptions of \textit{utility}, \textit{ease of use} and \textit{efficiency}, as well as the \textit{maturity of tools}. Companies with lower adoption of MDE often perceive tool maturity to be worse than those with higher adoption~\cite{Mohagheghi2013study}.

    \item \textbf{Versioning} is the ability to track and merge model changes, facilitating model maintenance, and is one of the main enablers of \textbf{Collaboration}, as stated in \cite{pietron2020collaborative}. MX in a team is highly dependent on combining different engineers' work and identifying differences between different versions of the same model.  This suggests that versioning is also an important factor of MX.
    In the literature, support for collaboration has consistently emerged as a desired attribute of modelling tools for practitioners\cite{Ozkaya2019Ltr,Badreddin2018trends, BordeleauLRST17,david2023collaborative,franzago2017collaborative}. In the more general field of SE, collaborative tools are found to positively impact productivity~\cite{hidayanto2014impact,ur2020use}.
    The lack of adequate version management support has been identified as a drawback of Model-Based Engineering (MBE) tools within the embedded systems domain~\cite{Grischa2014Embedded}. The main issue is that visual modelling languages are notoriously difficult to version -- because changes in the layout and changes in the semantics can be difficult to distinguish, and text-based diffing and merging tools that are integrated into tools like git or its frontends like GitHub or GitLab are not capable of diffing them visually at all. Specialised tools, such as EMF Compare\footnote{https://eclipse.dev/emf/compare/}, have improved over the years, but they still do not achieve the level of integration developers are already used to with code. 
    Recent work combining textual and visual representations of the same models attempts to address these issues~\cite{exelmans2022optimistic} fundamentally.

    \item [\textbf{Technical Capabilities}:] The capabilities resulting from the combination of the modelling language and tool to, e.g., reason about the modelled system's properties or generate downstream artefacts.


\end{description}

\paragraph{Non-technical factors}

\begin{description}
    \item [\textbf{Modeller-Intrinsic Motivation}:] Is driven by perceived benefits and positive emotions from modelling and leads modellers to use modelling. \textit{Intrinsic motivation} stands in contrast to extrinsic motivation, where the incentive originates from an external source rather than the inherent appeal of the task~\cite{Ryan2000Intrinsic}. Intrinsic motivation is a key predictor of developer experience in software development~\cite{Kuusinen2016Intrinsic}.

    \item [\textbf{Organisation-Intrinsic Motivation}:] Internal motivations within organisations significantly influence the adoption of modelling, driven by perceived benefits in productivity, quality, cost, and collective well-being. The literature highlights these benefits, particularly in the embedded software industry where efficiency and quality are key adoption motivators for MDE~\cite{Akdur2018Survey}. Additionally, factors like organisational culture, expertise, and internal advocacy significantly influence MDE adoption~\cite{Hutchinson2011Empirical}. 
    
    \item [\textbf{Organisation-Extrinsic Motivation}:] External factors influencing an organisation's adoption and approach to modelling include standards, regulations, tool availability, and customer demands. Adherence to regulations, especially in the embedded systems industry, is a significant driver for adopting MDE~\cite{Vogelsang2018Embedded}. 
    
    \item [\textbf{Training}:] Factors related to skills and knowledge in using modelling languages and tools. The importance of training for MDE adoption is highlighted \cite{Hutchinson2011Empirical}, with challenges in the \textit{insufficiency of training resources} \cite{kalantari2023unveiling} and \textit{developer training efforts}.
\end{description}

The working group believes that these factors are important aspects that affect MX and that understanding their relationships is crucial to understanding why modelling works (or does not) in an organisation or project and how to improve MX. To test this idea, we will apply these factors to the workflows already mentioned above. In these workflows, we captured typical well-known scenarios from practice and literature: informal modelling with and without digitally captured diagrams, architectural modelling in the classical model-based systems engineering sense in the automotive industry, and modelling as part of work on infrastructure-as-code. We believe that these scenarios are archetypes and that, by applying the factors to them, we can learn more about their applicability and gain insights into their relevance. 

As a result of this effort, we envision some possible future research questions:
\begin{itemize}
    \item Is there evidence that the proposed MX factors are relevant in practice? While we brought together significant experience in modelling from practice and research, our current understanding of the factors and their relationships should be validated in industrial settings with the help of practitioners. To achieve this, we plan to conduct an empirical study, possibly using a mixed-method design with interviews and a survey.
    \item Do existing modelling tools address the factors we have identified? On the one hand, answering this question will allow us to validate the relevance of the proposed factors. On the other hand, this mapping will also allow us to identify factors unsupported by a modelling approach or tool and are, therefore, candidates for further exploration.
\end{itemize}

A forthcoming paper that describes the factors in more detail and maps them to the archetypical workflows will be a first step towards answering these questions.

\section{Topic 2: Collaboration and MDE}
In the breakout group for Topic 2, we discussed the topic of collaboration and MDE, which may be parsed in at least three ways:
First, we can associate \emph{modelling together} to solve a problem, such as with interactive editors or version control systems trimmed for supporting modelling in teams.
Second, we can associate using models to \emph{express} collaborative development and, third, using models to \emph{leverage communication and, thus, collaboration} (between different roles in a software project).
While the first two options may provide several possibilities for research, in the sequel, we focus on the third direction of using models to support communication in software projects.

\subsection{Motivation}
In software projects, different roles and teams need to exchange information. 
Although models provide an abstraction layer based on which humans in different roles can communicate, models are not sufficiently used to \emph{share} information between roles. 
Since they provide abstractions which are trimmed to the purpose of the role, another role may not understand them. 
Thus, the missing capabilities of models for collaboration in software projects provoke lost, inconsistent, or untraceable information, such as requirements, design, and implementation decisions. 
The following example illustrates these issues.

\paragraph{Motivating Example}
Let us consider the transition from the requirements engineering to the architecture and design phase: 
requirements analysts should provide information on the results of their work to the software architect. 
Understanding which requirements must be satisfied and what they mean is crucial for project success. 
Consequently, it is essential that requirements analysts and software architects (and later on any other involved person in the process, such as developers, testers, etc.) form the same understanding of the requirements' meaning.

In case of a plan-driven process, 
a specification with use cases and use case diagrams results from the requirements engineering phase, representing 
 one possibility to model the elicited requirements. 
 Although use case diagrams are an established means of modelling requirements, it is debatable whether they suffice for software architects to understand for what software they should provide an architecture. 
 Instead, both roles communicate informally and verbally, which always introduces the risk of misunderstandings or lost information. 

Using persisted models provided by the requirements engineers, such as use case diagrams, can support establishing a shared understanding, but in practice, they tend to be neglected, particularly in communicating between project roles~\cite{petre2013umlInPractice,mussbacher2014relevance}.
\subsection{Characteristics of Models for Supporting Collaboration}

To solve the problem of refused communication between roles using models, we first need to examine which characteristics of models hinder or benefit communication and, thus, collaboration between different roles. 
In the following, we present key characteristics that we elicited in extensive focus group discussions and refined based on discussions with all participants.

\begin{figure*}
    \centering
    \includegraphics[width=0.7\textwidth]{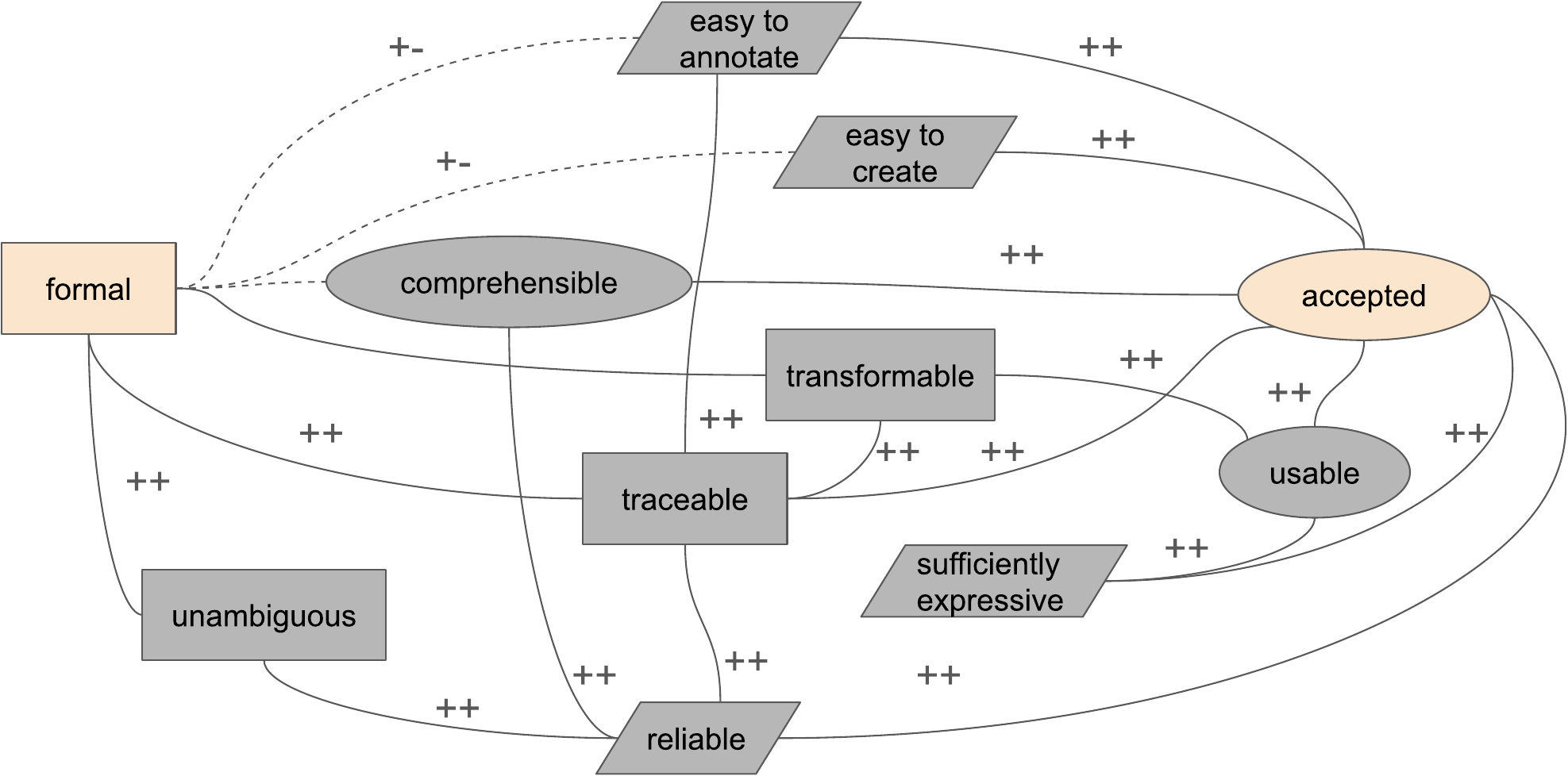}
    \caption{Characteristics of models to benefit communication in collaborations}
    \label{fig:collaborations_modelcriteria}
\end{figure*}
First, two main characteristics of models are decisive, the formality and the acceptance state, as depicted in orange in \autoref{fig:collaborations_modelcriteria}. 
We consider “acceptance” as the goal (necessary requirement) to be satisfied when we want a model to be used for communication. 
We identified the “formality” of the model as the independent variable which we can control. 
Although formality does not directly influence acceptance, several mediating factors exist that formality influences (e.g., the ease of use) and which, in turn, influence acceptance. 

Furthermore, the figure distinguishes the characteristics of a model that are technical-centred (visualised as rectangles) and human-centred (visualised as ellipses), and some that are shared by both (visualised as parallelograms). 
The association of an attribute to the human or the technical support for the model depends on the measurement. 
Human-centred characteristics completely depend on the human's assessment (and, hence, vary depending on the person interacting with the model) 
whereas one should be able to measure technical-centred attributes objectively. 
The figure presents relationships between the different characteristics with clear increasing or decreasing effect 
(“the more X, the more Y”, “the less X, the less Y”; denoted as “++”) or with contradicting effect 
(“the more X, the less Y” or “the less X, the more Y”; denoted as “+-” and vice versa, respectively). 
For example, the more formal a model is, the more complex it is to create, as it requires more effort to define it properly. 
Conversely, for instance, the more comprehensible a model is, the contained information is of higher reliability which, in turn, positively affects the acceptance of a model.

In summary, based on the figure we can conclude that 
the amount of formalism influences negatively the ease of annotation, creation, and comprehension. 
In contrast, the more formal a model is, the higher the likelihood it is unambiguous, transformable, traceable and, in turn, reliable and usable, positively affecting its acceptance. 

\subsection{Research Agenda}
As a result of eliciting these characteristics, we propose to pursue the following steps to optimise using models for effective communication and collaborations:
\begin{enumerate}
    \item {Investigate why different roles do not use models for communication across their field.} 
  
    \item {Empirically elicit characteristics in frequently used models which complicate or ease the communication across roles and stakeholders.} 
    \item Compare the characteristics with human needs to communicate
     provide guidelines that enhance existing models to facilitate communication across roles
    \item Derive metrics for measuring the quality of the enhanced models.
\end{enumerate}

By pursuing these steps, we will first derive insights from practice and research, which may require interdisciplinary models. 
Furthermore, the elicited characteristics allow us to define guidelines to optimally use models in communication.
Eventually providing metrics may increase the practical usage of models in communication across heterogeneous roles in software projects.

\section{Topic 3: Diversity and Inclusion in MDE}
In the breakout group for Topic 3, we critically discussed diversity and inclusion as related to MDE and formulated a research roadmap for the topic.
The diversity dimensions we consider are age, educational background, cultural background, gender, experience level, cognitive diversity (i.e., neurodiversity), and colour blindness of the modeller and the model end user.
We either have evidence suggesting differences concerning MDE for these dimensions, or we hypothesise a potential impact. 
For instance, cognitive diversity, user age, gender, or cultural background might all affect preferred working methods and problem-solving styles.
In turn, these preferences might align well or conflict with MDE tools, notations, and workflows.

\begin{figure*}[t]
    \centering
    \includegraphics[width=.6\textwidth]{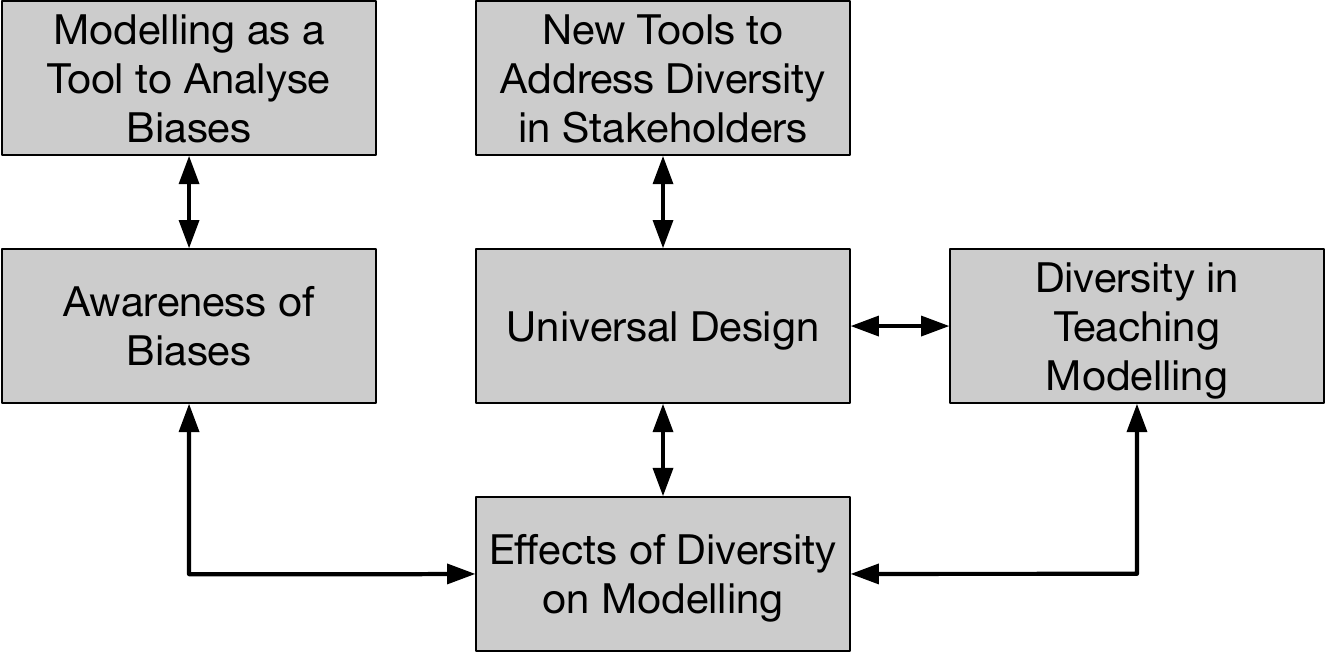}
    \caption{Overview of identified dimensions of research topics relevant for diversity in MDE.}
    \label{fig:dei-mde}
\end{figure*}

Several discussions and brainstorming sessions based on these diversity dimensions led us to six distinct research directions related to Diversity and Inclusion in MDE.
An overview of these directions and their connections is depicted in Figure~\ref{fig:dei-mde}.
In the following, we discuss each direction separately.

\subsection{Universal Design}
Universal Design is ``the design and composition of an environment so that it can be accessed, understood and used to the greatest extent possible by all people regardless of their age, size, ability or disability''\footnote{\url{https://universaldesign.ie/what-is-universal-design/}}.
We transfer this concept to MDE, proposing that a better understanding of what needs different diverse groups have will lead to improvements in MDE and related practice. Improving modelling tools, languages, techniques, and practices (MTLTPs) accordingly will ultimately benefit all people and the field, not only the specific group targeted.

Thus, the central goal of this research direction is to understand the needs for MTLTPs with respect to several diversity dimensions.
To address this goal, we need to investigate what aspects of modelling are affected by diversity concerns, and which barriers exist for diverse stakeholders.
This relates to tooling and visual aspects, e.g., modelling notations and layout, but also comprises cognitive aspects, such as preferred workflow, model comprehension, and information processing style.
Understanding these effects and the involved dimensions will allow for the creation of guidelines for better tool and practice design surrounding MDE that benefit all users.

\subsection{Awareness of Biases}
MTLTPs are purposefully designed but may contain biases, which raises concerns about their inherent subjectivity and their reflection on the modeller's background.

We hypothesise the presence of biases in MTLTPs toward specific diversity dimensions, highlighting the importance of studying these biases. The objective is to compile scenarios illustrating how MTLTPs might yield biased results, providing insights for refining design guidelines.

For instance, cultural biases may emerge in language design, affecting text length and layout preferences. In modelling videos from India, we observe that reference names and annotations tend to be longer than the common European approach, which might relate to low \& high context cultures~\cite{hall1976beyond}. Additionally, gender bias is noticeable in teaching materials through stereotypical representations, potentially contributing to inequities~\cite{medel2017eliminating}.

The overarching objective of this dimension is twofold: (1) Make biases explicit in models and (2) raise awareness to guide future MTLTP design.
This involves examining real-world examples of biases from diverse literature, assessing student assignments and critically evaluating teaching materials, papers on modelling, and content analysis of teaching videos.

\subsection{Effects of Diversity on Modelling}
The exploration of the effects of diversity on modelling centres on recognising that models are created, maintained, and interpreted by individuals with unique characteristics, influencing their interactions with models. 

Given the impact of diverse group compositions on communication and collaboration methods, the primary objective is to understand better how various diversity dimensions influence the modelling process within group dynamics, and subsequently affect the features of resulting models. 

Examples from other SE sub-fields, such as programming, demonstrate that individuals from different age groups~\cite{baltes2018towards} or genders~\cite{catolino2019gender} exhibit varied working approaches, stress coping mechanisms, and conflict resolution strategies, influencing the outcomes, such as code quality~\cite{catolino2019gender}.

The long-term goal involves creating awareness of expected differences, developing inclusive modelling tools, and enhancing accessibility in modelling explanations through comparative analysis of modelling discussions and decisions among diverse groups via participatory observations.

\subsection{Modelling as a Tool to Analyse Biases}
Models are a tool for abstraction. 
As such, we hypothesise that they may be more suitable for analysing and detecting certain biases that may be otherwise obfuscated in more detailed artefacts defined at lower abstraction levels, such as source code. 
This research direction aims to understand how we can leverage models to detect potential biases. This could be done by creating methods for systematically reasoning about diversity through models, or by using models in the context of participatory design to involve diverse stakeholders in the modelling process.

For example, reasoning about requirements models may facilitate detecting missing requirements that affect marginalised or underrepresented groups. In addition, models can also be created to help detect bias or avoid it~\cite{NUNES2023102108}.
Finally, models may be used to analyse potential biases in datasets built for machine learning, which could affect data collection, assumptions about the data, specific valid scenarios, or data quality. 

\subsection{Diversity in Teaching Modelling}
The effective teaching of MDE and modelling requires consideration of various diversity dimensions and their implications for modelling practices. Recognising and mitigating barriers to learning and adopting modelling techniques is crucial to fostering an inclusive and accessible educational environment.

To overcome these barriers, tailored teaching strategies and methodologies are crucial. For example, enhancing self-efficacy in female students has been shown to be essential for engaging them in computer science disciplines, including MDE \cite{vrieler2021computer}. Furthermore, the absence of timely feedback during the learning process can result in frustration, adversely affecting students' willingness to embrace and learn modelling. Students on the autistic spectrum may encounter difficulties with free-hand modelling exercises, owing to a need for clear processes and well-defined rules \cite{baron2005autism}.

Addressing the needs of diverse groups during teaching can potentially reduce barriers to adoption. Key questions for exploration include how teaching methods and tools impact different groups, identifying teaching strategies that may conflict with diverse needs, and understanding how to mitigate conflicts arising from diversity within student groups to ensure a positive learning experience. The long-term goal is to develop and promote inclusive MDE teaching practices through collaborative efforts involving representatives of diverse groups. Evaluating developed teaching methods on a larger scale, including international perspectives, will be essential to gauge effectiveness and ensure inclusion.

\subsection{New Tools to Address Diversity in Stakeholders}
Enabling more diversity in modelling also asks for a technical view on MDE. Tools and languages, including Domain-Specific Languages (DSLs), provide the basis for modelling. Thus, the tools and modelling languages in use need to be designed to be inclusive of the needs of diverse users. If tools are not designed according to the requirements of a diverse group of users, they introduce additional obstacles to the modelling process \cite{NUNES2023102108}.

We hypothesise that the literature already provides substantial approaches that offer potential solutions. For example, the dedicated creation of DSLs or Blended Modelling~\cite{david2023blended} environments for specific stakeholders or domain experts can be further used to address stakeholder diversity as well. By offering a variety of concrete syntaxes related to the same abstract syntax, different groups can choose the concrete syntax that successfully addresses their particular needs.

To succeed in the goal of providing tools and modelling languages that are inclusive of the needs of diverse users, we propose studying  the following high-level questions:
\begin{itemize}
    \item How can modelling tools successfully address the needs of diverse groups?
    \item How can tools address collaborative modelling when users prefer different syntaxes/representations? 
    \item How can blended modelling languages or DSLs be designed for the use of diverse groups?
\end{itemize}




The long-term goal of the research on this issue should be to develop guidelines limiting the barriers to using modelling tools. These guidelines can also help evaluate newly developed tools on their inclusiveness for diverse groups.

\section{Topic 4: Modelling Human Factors}

In the breakout group for Topic 4, we discussed that when modelling human factors, many possible aspects could be considered. We strived for an interdisciplinary topic between software engineering and psychology to find a concrete aspect that we could work on during the seminar.

\subsection{Motivation}
Over the last few years, the awareness of human issues and individual human aspects in SE has grown. There are currently specialised development processes (such as Human-Centred Design) that emphasise human involvement. Therefore, we discussed what other aspects of humans could be integrated into our existing approaches. 
Previous psychology research has aimed to model or represent different aspects influencing human behaviour. Humans are complex beings, which makes it difficult to create holistic models. However, individual human aspects can be modelled currently with sufficient significance. 

\subsection{Human Values}
\begin{figure}[tb]
    \centering
    \includegraphics[width=0.9\linewidth]{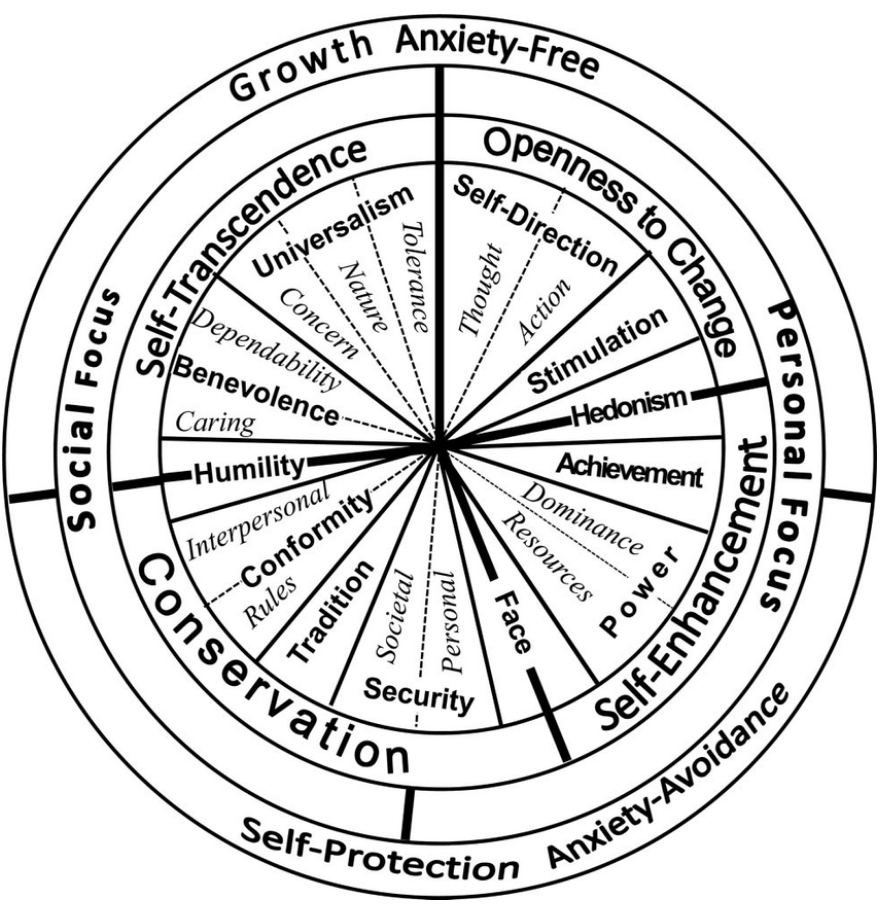}
    \caption{Circular model of Schwartz's theory of basic human values. (Source: Schwartz et al.~\cite{schwartz2012overview})}
    \label{fig:schwartz-model}
\end{figure}

One of these aspects are human values, as proposed within Schwartz's theory of basic human values~\cite{schwartz2012overview}. These values are modelled in relation to each other, as depicted in Figure~\ref{fig:schwartz-model}, where nearby values are related, whereas bipolar values are conflicting.

From the SE perspective, human values are crucial.
Human values of end users play an important role in software projects: values such as trustworthiness, fairness, etc., are becoming increasingly important, and companies therefore want to be associated with such positive values. Until now, human values have hardly been explicitly taken into account in the development process (e.g., as non-functional requirements). 

\subsection{Process}
We envision a human value-centric software process that deliberately consists largely of familiar building blocks but puts them together in a new way. This is because the human values of end users are often vaguely formulated at the beginning and have different meanings for different people. Therefore, in the development process, these values must be captured during the requirements elicitation and passed on to the stakeholders and developers.
Traceability is crucial so that human values do not remain an abstract concept but are linked to design and implementation artefacts.

\subsection{Novelty}
Building on existing concepts such as user involvement and personas, our envisioned approach aims to transcend individual human desires to focus on emerging human values, using the Schwartz model as an example. It provides a pragmatic technique for constructing personas as a collection of human aspects, focusing on human values. The approach suggests a systematic progression from high-level needs to decisions that guide the implementation of value-related requirements. It proposes a systematic framework suitable for ordinary software organisations. The approach is not limited to human-centred design professionals but also considers developers, business owners, and other stakeholders to address the challenges related to the ambiguity of human values. Additionally, we are developing important research questions to enhance and evaluate this approach effectively.

\subsection{Directions}
After the seminar, we worked on a systematic approach to represent human values and ensure traceability.
In the near future, we will publish a separate research article that describes the detailed approach.

\section{Topic 5: Teaching Human-aware MDE}
\label{sec:topic5}
Models or sketches for thinking are universally beneficial due to their ability to clarify complex ideas and improve understanding. In the realm of software and systems engineering, models serve as essential tools for simplifying complex systems, enabling a shared language among stakeholders, enhancing collaboration efficiency, and improving decision-making processes. Effective modelling, especially in open-ended problems, necessitates structured thinking and a clear purpose, ensuring that all aspects of the system are considered and appropriately addressed.

Teaching human-aware MDE goes beyond the dissemination of tools and techniques; it involves nurturing a mindset geared towards prioritisation, in-depth analysis, and strategic approaches to complex issues. This section discusses the educational frameworks and pedagogical strategies essential for cultivating such a mindset. 

\subsection{Pedagogical Challenges}

\paragraph{The Confidence in Modelling Skills}
A key challenge in teaching modelling skills is students' lack of confidence in their modelling abilities. This often stems from a fear of making mistakes or not having complete knowledge of the modelled domain, leading to scepticism towards their models. This scepticism towards models can be significantly mitigated by teaching structured thinking, which provides a clear, strategic framework for tackling modelling problems.
 

A significant problem faced by computer science students is the premature focus on implementation during initial designs, which inhibits the development and utilisation of abstraction skills. This focus on implementation is a manifestation of difficulties in separating problem space and solution space~\cite{stikkolorum2022studies}. 

It is crucial for educational approaches to focus on exploring the problem space and abstract thinking. Abstraction in modelling involves the process of identifying and focusing on the essential aspects of a real-world problem or system while disregarding the less significant details. This selective focus allows modellers to create simplified representations (models) of complex systems that retain only the features pertinent to the problem at hand. The goal is to make the problem more manageable and the system more understandable without oversimplifying it to the point of losing essential characteristics.

In contrast to the typical computer science view where complexity is hidden in nested levels of abstraction, abstraction in modelling is driven by the need to explore and understand the problem space. This involves:

\begin{description}
    \item [Problem Space Exploration:] Understanding the various elements, relationships, and processes that constitute the system. This step is crucial for identifying what should be included in the model and what can be left out.
    \item [Purpose-relevant Information Extraction:] Determining which aspects of the system are relevant to the goals of the modelling effort. This means different models might be created from the same system for different purposes, each abstracting different facets or details based on what is most relevant to the problem or question being addressed.
    \item [Creation of Usable Models:] Developing a representation of the system that is both simpler than the reality but still retains all necessary detail to be effective for solving the problem or answering the question at hand. This involves a balance between simplification and accuracy, ensuring that the model is both tractable and valid within the context of its intended use.
\end{description}

Abstraction, therefore, is a critical skill in modelling because it allows the modeller to construct meaningful and usable models without becoming overwhelmed by the complexities of the real world. It involves a deliberate process of filtering out noise and focusing on the essence of the problem, which requires a deep understanding of both the system being modelled and the context in which the model will be applied. This skill enables modellers to navigate the vast problem spaces they encounter and develop effective solutions and strategies. 
This will better prepare students to contribute effectively in the rapidly evolving technological landscape. Abstraction skills are difficult to teach as they require balancing detail and generalisation.

\paragraph{The Crucial Role of Feedback}

Feedback is an indispensable component in the learning process, as it supports learning and builds confidence. The lack of relevant and constructive feedback contributes to the challenges students and teachers face. Furthermore, the previously mentioned lack of confidence of students in their modelling abilities is often exacerbated by the absence of relevant and constructive feedback on the quality of their models.

However, providing quality feedback on models presents unique challenges. For example:

\begin{description}
    \item [Assessing completeness:] In complex domains, it can be difficult to determine if a model comprehensively covers all relevant concepts.
    \item [Evaluating correctness:] The correctness of a model goes beyond just accurately representing individual concepts; it also involves the relationships between these concepts.
    \item [Navigating multiple valid models:] For a given problem within a domain, there can be several different yet equally valid models. This complicates the feedback process as there is no single ``right'' model against which students' work can be benchmarked.
    \item [Understanding model quality:] There is no tangible, universally accepted definition of model quality; hence providing feedback on model quality can be abstract and challenging for students to grasp.
\end{description}

Addressing these challenges necessitates a nuanced approach to feedback that recognises the complexities involved in modelling. 

\paragraph{The Embracing of and Motivation for Modelling}

Many students find it difficult to appreciate the significance of modelling. Modelling can be seen as an overwhelming abstract and annoying task, being in the way to start programming. 
This view can lead to a lack of interest in mastering modelling techniques and recognising their application across various fields.

The true benefit of modelling manifests itself in collaborative, iterative, and long-term projects. However,  these situations are rarely utilised in education. For instance, while group projects are common, they rarely focus explicitly on fostering collaborative modelling skills. Additionally, the typical duration of university courses, spanning from a few weeks to a semester, limits the depth of engagement students can have with complex modelling projects, thus restricting the potential for meaningful teamwork and iterative learning.

Finally, our observations suggest that the ability of students to read and comprehend existing models is underdeveloped. This deficiency stems from educational practices that focus on creating new models, neglecting the equally important skill of model interpretation. Students seldom receive assignments that require them to analyse and understand existing models, which leads to a lack of familiarity and confidence in dealing with pre-established frameworks. Furthermore, this leads students to adopt a skewed understanding of modelling. This gap undermines their confidence in modelling and impairs their ability to effectively contribute to ongoing projects or adapt existing models to new contexts.


\subsection{Addressing the Challenges}
Addressing the challenges outlined above requires an integrated approach to teaching modelling that highlights practical benefits, encourages active collaboration, incorporates activities and assignments that focus on the critical analysis of existing models, and demystifies the various techniques and notations used. 

\paragraph{Strategies for Assessing Modelling Skills}


The challenge in teaching modelling skills is finding an equilibrium between the need for tangible, assessable outcomes and maintaining the holistic nature of of modelling, which is critical for students' comprehension and application in real-world scenarios. Educators must navigate this balance to cultivate both technical competence and an appreciation for the value of modelling in the broader interdisciplinary educational context.

To address this challenge, educators should consider implementing a multifaceted approach that includes incremental learning and constructive feedback on modelling assignments, structured thinking exercises that promote abstract reasoning and problem decomposition, and practical training through real-life case studies and examples. This approach, including critical essays, analysis papers, Socratic discussion seminars, peer reviews and case studies, fosters a more profound understanding of the importance and applicability of modelling in various contexts. 

\paragraph{Strategies for Delivering Constructive and Relevant Feedback}

It can be beneficial to follow a structured guideline when assessing student models to elevate the quality of students' models and provide better feedback. A key element of this strategy involves differentiating feedback between syntax and semantics. By distinguishing between these two dimensions, educators can provide more targeted feedback that addresses student models' form and substance. Moreover, feedback should be positive and constructive, highlighting what students have done well and pointing out areas for improvement. This approach encourages learning by reinforcing successful strategies and providing clear, actionable suggestions for addressing weaknesses. Such feedback fosters a learning environment that values growth and clear, targeted improvement.

Feedback must be specific and refer to concrete instances within the model to be effective. For example, rather than making a general statement about the model's organisation, an instructor might point out how the placement of a particular element enhances the model's clarity and coherence. Additionally, feedback should clearly differentiate between the problem space and the solution space. This distinction helps students understand whether issues arise from their understanding of the problem or the resolution methods they have chosen to solve.

When measuring the quality of student models, visual aesthetics and the layout of diagrams can serve as an initial proxy for model quality, as they often reflect the student's attention to detail and organisational skills. However, one should be careful not to overemphasise form over function. The naming of model elements is also critical, as names should be descriptive, concise, and appropriate to the level of abstraction while also conforming to established conventions. 

Finally, an attempt at completeness should be made by comparing the list of concepts in the problem description with those present in the model, identifying any missing or unnecessary elements. 
Educators can significantly enhance students' modelling skills through such thorough and thoughtful feedback, providing them with the tools and understanding necessary to develop high-quality, meaningful models.

\paragraph{Strategies for Teaching Structured Thinking}
Structured thinking is a foundational skill for modellers, fostering confidence in their models and underlining the relevance of modelling. It enables the breakdown of complex problems into manageable parts, facilitating clear analysis. Teaching structured thinking involves introducing students to frameworks and methodologies that guide the systematic exploration of problems and the development of coherent, logical models. Techniques such as mind mapping, decision trees, and logical frameworks can be instrumental in this process.

This educational approach should focus on iterative learning, where students iteratively refine their critical and strategic thinking abilities. 
Encouraging reflective practice, where students evaluate their thought processes and modelling decisions, can significantly enhance their understanding and application of structured thinking in modelling. 
Exercises designed to simulate real-world scenarios can provide practical experience in applying structured thinking, thereby enhancing students’ confidence in their modelling abilities and the perceived value of modelling.

\section{Discussion}\label{sec12}
\begin{figure*}[t]
    \centering
    \includegraphics[width=.7\textwidth]{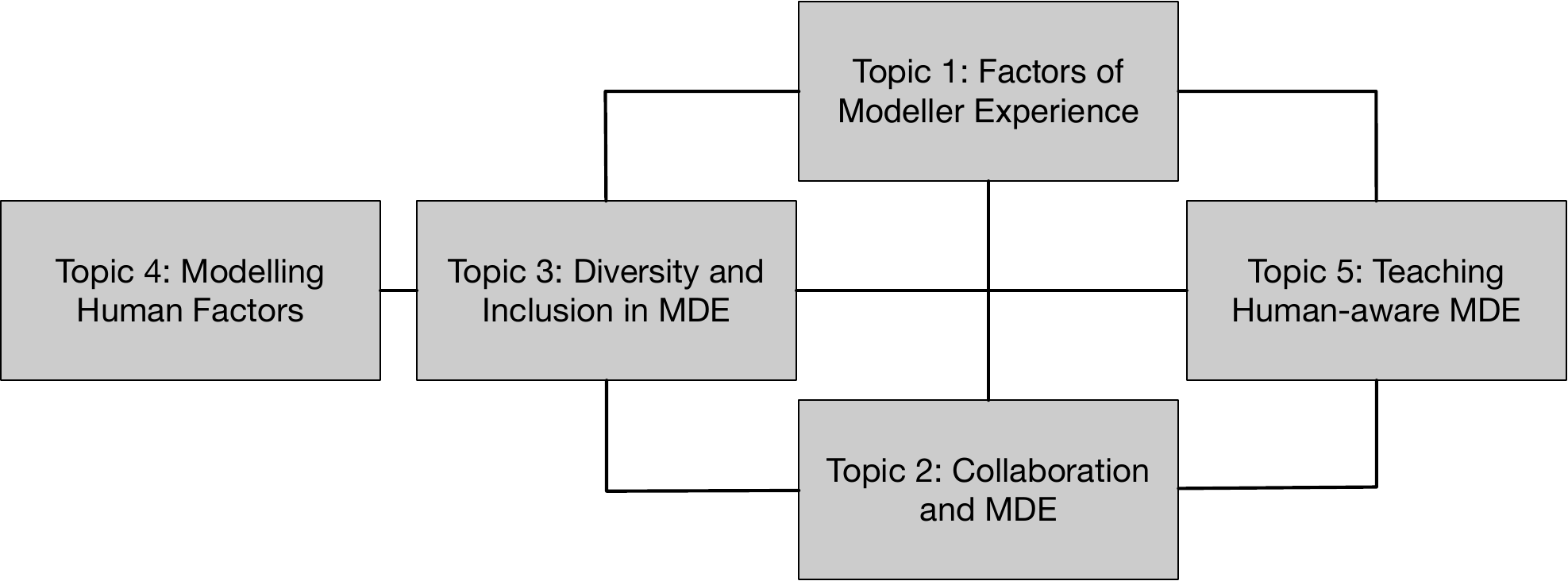}
    \caption{Overview of the seminar topics.}
    \label{fig:topics}
\end{figure*}

The topics summarised in Sections~\ref{sec:topic1} to \ref{sec:topic5} outline various topics, goals, and research questions for future work.
The five topics cover specific aspects concerning those topics in depth.
However, there are various connection points between the five topics that are worth highlighting.
These connections are sketched in Figure~\ref{fig:topics}.

Overall, we observe that topics 1, 2, 3, and 5 are well-connected.
Topic 1, \emph{Factors of Modeller Experience}, provides a broad frame of technical and non-technical factors that can affect modeller experience.
Topic 2, \emph{Collaboration and MDE}, and Topic 3, \emph{Diversity and Inclusion in MDE}, consider specific cases of such factors, i.e., diversity-related factors and communication/collaboration-related factors.
These specific cases might give rise to additional MX-related factors, or moderate their effect.
Jointly, the three topics promote the message that considering technical factors alone is insufficient for successful use and adoption of MDE, and to understand how and why MDE is applied in practice.
These observations lead us to the following proposition:
\greybox{
\textbf{Proposition 1:} We propose that modelling is not an objective process, but various technical and non-technical factors affect the individual modelling experience. The community needs to (1) investigate what these factors are, and (2) investigate to what extent the factors affect modelling practice and adoption.
}

Topics 1 to 3 also all discuss that specific contexts need to be distinguished and studied separately, which is introduced as archetype workflows in \emph{Factors of Modeller Experience}.
While named differently, specific examples of such workflows are covered in \emph{Diversity and Inclusion in MDE} with respect to diversity \& inclusion, and in \emph{Collaboration and MDE} with respect to collaborative aspects of MDE.
Collecting such archetype workflows could be an important goal for the MDE community as a whole, as it would allow to better reason about the (external) validity of various empirical studies.
We summarise this as follows:
\greybox{
\textbf{Proposition 2:} We propose that various different contexts or workflows need to be considered in MDE research to allow for more precise conclusions to be drawn. The community needs to collect and validate representative workflows of the application of MDE in practice.
}
Jointly addressing Propositions 1 and 2 could help the MDE community increase the validity of research results, e.g., in reasoning about success factors of MDE-related interventions in industry.

%
Topic 5, \emph{Teaching Human-aware MDE}, connects naturally to Topics 1 to 3 as, ultimately, students (and future engineers) need to be taught the complexity of modelling beyond purely syntactical and technical factors.
On the one hand, this necessitates that educators are aware of relevant human factors and dimensions that affect MX so that MX does not negatively affect the classroom, ultimately leading to a lower acceptance of MDE.
On the other hand, it also requires highlighting these factors to the student population to ensure transfer to and awareness in practice.
We summarise these observations as follows.
\greybox{
\textbf{Proposition 3:} Education in MDE needs to mirror research efforts in the sense that educators (1) need to understand the effect of human factors on modeller (and student) experience, and (2) need to convey the importance of these human factors to the student population, so that knowledge transfer to practice takes place.
}

Topic 4, \emph{Modelling Human Factors}, is somewhat disconnected in contrast to the close coupling of the remaining topics.
This group takes a different approach, namely how models can be used to analyse or consider human factors in SE.
Topic 3, \emph{Diversity and Inclusion in MDE}, covers this angle as well in two sub-points, i.e., that models can be a means to better understand diversity-related biases, and that existing contributions in the MDE community can be used to better support diverse needs among modellers and consumers of models in practice.
This angle can be compared to previous discussions on how knowledge in the MDE community can serve other research fields, e.g., for climate modelling \cite{easterbrook2015modelling,blair2016grand}.
However, in these previous cases, authors have advocated applying MDE-related knowledge to better understand phenomena in the scientific disciplines of engineering and natural sciences.
Instead, we raise the point that the many contributions of the MDE community and their technical sophistication can and should be used for human factors.
Overall, we propose the following.
\greybox{
\textbf{Proposition 4:} Human factors are increasingly important in SE research and practice. Therefore, the community needs to apply their MDE-related knowledge and experience to models of human factors in SE.
}

In addition to the research goals and questions suggested by the individual groups, and the propositions made above, connections between individual topics can lead to further interesting avenues for future work.
For instance, Topic 2, \emph{Collaboration and MDE} naturally connects to several of the other topics for issues concerning collaboration or teamwork.
Specifically, Modeller Experience, as covered in Topic 1, might vary from individual model use in cases of collaborative modelling.
\emph{Versioning} is highlighted as a technical factor contributing to MX and essential for successful collaborative MDE.
However, further technical and non-technical factors could similarly affect MX in collaborative settings.
Finally, \emph{Collaboration and MDE} has a joint connection to Topic 3, \emph{Diversity and Inclusion in MDE}.
Collaborative modelling in a diverse group of people can lead to diverse, and potentially conflicting viewpoints, opinions, and preferences.
This influence of diversity needs to be better understood to enable successful collaborative modelling practices.
Similarly, in education, collaborative modelling in diverse groups needs to be learned and practised to enable university graduates to deal with this diversity in practice.
Topic 1, \emph{Factors of Modeller Experience}, has an additional connection to \emph{Diversity and Inclusion in MDE}.
That is, several non-technical factors related to MX might be affected by the diversity of the modeller(s) or the model's users.
For example, preferences on factors such as layout, tooling, or amount of information contained in a model can potentially affect MX.
At a higher level of abstraction, it is worth considering whether diversity-related topics add additional factors not considered in \emph{Factors of Modeller Experience}, such as cultural factors, or if they are simply moderators to the listed factors.
In summary, we therefore propose the following.
\greybox{
\textbf{Proposition 5:} In addition to the points raised in Proposition 1 to 4, the community needs to pay considerable attention to how collaboration and group diversity affects modelling.
}

\section{Conclusion}\label{sec13}
This paper summarises the discussions we, a group of 26 researchers and practitioners connected to MDE and human factors research, had during a one-week seminar at Schloss Dagstuhl.
We focused on five topics which we elicited jointly at the start of the seminar.
In our opinion, these five topics require urgent attention in the MDE research community.
We formulated research goals, questions, and propositions that support directing future initiatives towards an MDE community that is aware of and supportive of human factors and values.

The topics raised represent the collective opinions and knowledge of the various experts present at the seminar, and are often not evidence-based.
Naturally, this means that others might disagree with our positions, and empirical evidence might prove us wrong in various ways in the future.
We encourage the community to engage with us in constructive discussions on our positions and suggestions.

Finally, several suggested topics did not receive enough attention during the seminar.
The list of goals, outcomes, and topics we presented in Section~\ref{sec:topicOverview} of this report can serve as an inspiration to explore further directions in addition to the ones we proposed. 

\backmatter

\bmhead{Acknowledgments}
We thank Schloss Dagstuhl for the opportunity to conduct the workshop and the GI (Gesellschaft für Informatik) for supporting it. 

\appendix

\section{Appendix: Goals and Setup of the GI-Dagstuhl Seminar}\label{sec:seminar}

The results presented in this paper emerge from discussions at a GI-Dagstuhl seminar\footnote{\url{https://www.dagstuhl.de/en/seminars/gi-dagstuhl-seminars}}.
These seminars are similar to the well-known traditional Dagstuhl seminars\footnote{\url{https://www.dagstuhl.de/en/seminars/dagstuhl-seminars}} but cater towards early-career researchers, such as PhD students and postdocs.

The primary goals of the seminar were (1) to establish closer collaborations and shape future goals in human factors in MDE and (2) to broaden the perspective of the seminar topic through insights into related research areas.
To address (1), we invited early-career researchers who study human factors in software modelling through contacts and existing publications in the HuFaMo workshop series, held as a part of the Models conference every year.
Additionally, we invited a few selected senior researchers in this community.
To address (2), we invited participants who conduct research not directly related to MDE but related to the human factors seminar theme, such as psychology or human-centred software engineering.
Finally, connected to both goals, we invited experienced industrial participants interested in MDE, to anchor discussions in industrial needs.

\subsection{Participants}
The aforementioned goals led to a diverse group of 26 participants, with 11 women and 15 men, 3 participants from industry, another 2 who used to work in industry, 9 participants before finishing their PhD, and 5 senior participants. Given the nature of Dagstuhl seminars, most of the participants came from Germany (15), followed by Iceland (2), Portugal (2), and Sweden (2). One participant each came from Belgium, Canada, Switzerland, the Netherlands, and Spain.  

\subsection{Seminar Structure}
After a first unofficial get-together on Sunday, we started the seminar early on Monday, 19th November 2023. We ended after lunch on Friday, 24th November 2023, spending a total of 4.5 days on in-depth discussion.
Throughout the workshop, we followed several of the practices recommended as a part of Liberating Structures\footnote{\url{https://www.liberatingstructures.com/}}, i.e., simple practices for structuring workshops and discussions engagingly and inclusively.

Monday was used to brainstorm seminar outcomes and topics.
From Tuesday until Friday, we discussed the five most popular topics in breakout groups.
Finally, we recorded results as a joint working document from Thursday onwards.
In parallel to these activities, we had eight selected presentations from junior participants and one selected presentation on psychological theories from an established researcher.

\subsection{Reflections on the Seminar Format}
During one week, the author group discussed and reflected on human factors in MDE in a GI-Dagstuhl Seminar.
The format we chose is akin to similar seminars, e.g., regular Dagstuhl Seminars, Lorentz Center Workshops\footnote{\url{https://www.lorentzcenter.nl/what-is-a-lorentz-center-workshop.html}}, or NII Shonan Meetings\footnote{\url{https://shonan.nii.ac.jp/proposal/guidelines/}}.
However, the seminar was aimed primarily towards early researchers, and we, on purpose, invited people with non-MDE backgrounds, i.e., conducting research in education and psychology.
Additionally, we invited several practitioners with a background in MDE research.
This diverse group composition led to many interesting and novel discussions, as participants were more likely to think ``out of the box'', i.e., make connections to non-MDE research topics or industrial practice.
Similarly, the juniority of the group as a whole might indeed have created more novel discussions, as many participants were not as mechanised in their thoughts on what MDE research is and should be.
Hence, we believe the diverse group composition was a success and will consider this for future seminars with similar aims.

%
%
%
%
%

%

%


\bibliographystyle{sn-nature}
\bibliography{sn-bibliography}

\begin{thebibliography}{10}
\expandafter\ifx\csname url\endcsname\relax
  \def\url#1{\burl{#1}}\fi
\expandafter\ifx\csname urlprefix\endcsname\relax\def\urlprefix{URL }\fi
\providecommand{\bibinfo}[2]{#2}
\providecommand{\eprint}[2][]{\url{#2}}
\providecommand{\doi}[1]{\url{https://doi.org/#1}}
\bibcommenthead

\bibitem{jolak2020software}
\bibinfo{author}{Jolak, R.} \emph{et~al.}
\newblock \bibinfo{title}{Software engineering whispers: The effect of textual
  vs. graphical software design descriptions on software design communication}.
\newblock \emph{\bibinfo{journal}{Empirical software engineering}}
  \textbf{\bibinfo{volume}{25}}, \bibinfo{pages}{4427--4471}
  (\bibinfo{year}{2020}).

\bibitem{ho2017practices}
\bibinfo{author}{Ho-Quang, T.}, \bibinfo{author}{Hebig, R.},
  \bibinfo{author}{Robles, G.}, \bibinfo{author}{Chaudron, M.~R.} \&
  \bibinfo{author}{Fernandez, M.~A.}
\newblock \bibinfo{editor}{Juristo, N.} \& \bibinfo{editor}{Shepherd, D.} (eds)
  \emph{\bibinfo{title}{Practices and perceptions of uml use in open source
  projects}}.
\newblock (eds \bibinfo{editor}{Juristo, N.} \& \bibinfo{editor}{Shepherd, D.})
  \emph{\bibinfo{booktitle}{2017 IEEE/ACM 39th International Conference on
  Software Engineering: Software Engineering in Practice Track (ICSE-SEIP)}},
  \bibinfo{pages}{203--212} (\bibinfo{organization}{IEEE},
  \bibinfo{year}{2017}).

\bibitem{gotz2021claimed}
\bibinfo{author}{G{\"o}tz, S.}, \bibinfo{author}{Tichy, M.} \&
  \bibinfo{author}{Groner, R.}
\newblock \bibinfo{title}{Claimed advantages and disadvantages of (dedicated)
  model transformation languages: a systematic literature review}.
\newblock \emph{\bibinfo{journal}{Software and Systems Modeling}}
  \textbf{\bibinfo{volume}{20}}, \bibinfo{pages}{469--503}
  (\bibinfo{year}{2021}).

\bibitem{liebel2019use}
\bibinfo{author}{Liebel, G.}, \bibinfo{author}{Tichy, M.} \&
  \bibinfo{author}{Knauss, E.}
\newblock \bibinfo{title}{Use, potential, and showstoppers of models in
  automotive requirements engineering}.
\newblock \emph{\bibinfo{journal}{Software \& Systems Modeling}}
  \textbf{\bibinfo{volume}{18}}, \bibinfo{pages}{2587--2607}
  (\bibinfo{year}{2019}).

\bibitem{hoppner2022advantages}
\bibinfo{author}{H{\"o}ppner, S.}, \bibinfo{author}{Haas, Y.},
  \bibinfo{author}{Tichy, M.} \& \bibinfo{author}{Juhnke, K.}
\newblock \bibinfo{title}{Advantages and disadvantages of (dedicated) model
  transformation languages: A qualitative interview study}.
\newblock \emph{\bibinfo{journal}{Empirical Software Engineering}}
  \textbf{\bibinfo{volume}{27}}, \bibinfo{pages}{159} (\bibinfo{year}{2022}).

\bibitem{France2007Seminal}
\bibinfo{author}{France, R.} \& \bibinfo{author}{Rumpe, B.}
\newblock \bibinfo{editor}{Briand, L.~C.} \& \bibinfo{editor}{Wolf, A.~L.}
  (eds) \emph{\bibinfo{title}{Model-driven development of complex software: A
  research roadmap}}.
\newblock (eds \bibinfo{editor}{Briand, L.~C.} \& \bibinfo{editor}{Wolf,
  A.~L.}) \emph{\bibinfo{booktitle}{Future of Software Engineering (FOSE
  '07)}}, \bibinfo{pages}{37--54} (\bibinfo{year}{2007}).

\bibitem{kalantari2022slr}
\bibinfo{author}{Kalantari, R.} \& \bibinfo{author}{Lethbridge, T.~C.}
\newblock \bibinfo{title}{Characterizing ux evaluation in software modeling
  tools: A literature review}.
\newblock \emph{\bibinfo{journal}{IEEE Access}} \textbf{\bibinfo{volume}{10}},
  \bibinfo{pages}{131509--131527} (\bibinfo{year}{2022}).

\bibitem{evans2014domain}
\bibinfo{author}{Evans, E.}
\newblock \emph{\bibinfo{title}{Domain-Driven Design Reference: Definitions and
  Pattern Summaries}}  (\bibinfo{publisher}{Dog Ear Publishing},
  \bibinfo{address}{Alaska, USA}, \bibinfo{year}{2014}).

\bibitem{gonzalez2013defining}
\bibinfo{author}{Gonz{\'a}lez-Huerta, J.}, \bibinfo{author}{Insfr{\'a}n, E.} \&
  \bibinfo{author}{Abrah{\~a}o, S.}
\newblock \bibinfo{editor}{Moreira, A.}, \bibinfo{editor}{Schätz, B.},
  \bibinfo{editor}{Gray, J.}, \bibinfo{editor}{Vallecillo, A.} \&
  \bibinfo{editor}{Clarke, P.} (eds) \emph{\bibinfo{title}{Defining and
  validating a multimodel approach for product architecture derivation and
  improvement}}.
\newblock (eds \bibinfo{editor}{Moreira, A.}, \bibinfo{editor}{Schätz, B.},
  \bibinfo{editor}{Gray, J.}, \bibinfo{editor}{Vallecillo, A.} \&
  \bibinfo{editor}{Clarke, P.}) \emph{\bibinfo{booktitle}{Model-Driven
  Engineering Languages and Systems: 16th International Conference, MODELS
  2013, Miami, FL, USA, September 29--October 4, 2013. Proceedings 16}},
  \bibinfo{pages}{388--404} (\bibinfo{organization}{Springer},
  \bibinfo{year}{2013}).

\bibitem{Whittle2015Taxonomy}
\bibinfo{author}{Whittle, J.}, \bibinfo{author}{Hutchinson, J.},
  \bibinfo{author}{Rouncefield, M.}, \bibinfo{author}{Burden, H.} \&
  \bibinfo{author}{Heldal, R.}
\newblock \bibinfo{title}{A taxonomy of tool-related issues affecting the
  adoption of model-driven engineering}.
\newblock \emph{\bibinfo{journal}{Software \& Systems Modeling}}
  \textbf{\bibinfo{volume}{16}} (\bibinfo{year}{2015}).

\bibitem{iso9241-210}
\bibinfo{author}{{Technical Committee ISO/TC 159/SC 4}}.
\newblock \bibinfo{title}{{ISO 9241-210:2019 Ergonomics of human-system
  interaction -- Part 210: Human-centred design for interactive systems}}.
\newblock \bibinfo{type}{Tech. Rep.}, \bibinfo{institution}{{International
  Organization for Standardization (ISO)}} (\bibinfo{year}{2019}).
\newblock \urlprefix\url{https://www.iso.org/standard/77520.html}.

\bibitem{Mohagheghi2013study}
\bibinfo{author}{Mohagheghi, P.}, \bibinfo{author}{Gilani, W.},
  \bibinfo{author}{Stefanescu, A.} \& \bibinfo{author}{Fernandez, M.~A.}
\newblock \bibinfo{title}{An empirical study of the state of the practice and
  acceptance of model-driven engineering in four industrial cases}.
\newblock \emph{\bibinfo{journal}{Empirical Software Engineering}}
  \textbf{\bibinfo{volume}{18}}, \bibinfo{pages}{89--116}
  (\bibinfo{year}{2013}).

\bibitem{pietron2020collaborative}
\bibinfo{author}{Pietron, J.}
\newblock \bibinfo{editor}{Guerra, E.} \& \bibinfo{editor}{Iovino, L.} (eds)
  \emph{\bibinfo{title}{Enhancing collaborative modeling}}.
\newblock (eds \bibinfo{editor}{Guerra, E.} \& \bibinfo{editor}{Iovino, L.})
  \emph{\bibinfo{booktitle}{Proceedings of the 23rd ACM/IEEE International
  Conference on Model Driven Engineering Languages and Systems: Companion
  Proceedings}} (\bibinfo{publisher}{Association for Computing Machinery},
  \bibinfo{year}{2020}).

\bibitem{Ozkaya2019Ltr}
\bibinfo{author}{Ozkaya, M.}
\newblock \bibinfo{title}{Are the uml modeling tools powerful enough for
  practitioners? a literature review}.
\newblock \emph{\bibinfo{journal}{IET Software}} \textbf{\bibinfo{volume}{13}}
  (\bibinfo{year}{2019}).

\bibitem{Badreddin2018trends}
\bibinfo{author}{Badreddin, O.}, \bibinfo{author}{Khandoker, R.},
  \bibinfo{author}{Forward, A.}, \bibinfo{author}{Masmali, O.} \&
  \bibinfo{author}{Lethbridge, T.~C.}
\newblock \bibinfo{editor}{Wasowski, A.}, \bibinfo{editor}{Paige, R.} \&
  \bibinfo{editor}{Øystein Haugen} (eds) \emph{\bibinfo{title}{A decade of
  software design and modeling: A survey to uncover trends of the practice}}.
\newblock (eds \bibinfo{editor}{Wasowski, A.}, \bibinfo{editor}{Paige, R.} \&
  \bibinfo{editor}{Øystein Haugen}) , MODELS '18, \bibinfo{pages}{245–255}
  (\bibinfo{publisher}{Association for Computing Machinery},
  \bibinfo{address}{New York, NY, USA}, \bibinfo{year}{2018}).

\bibitem{BordeleauLRST17}
\bibinfo{author}{Bordeleau, F.}, \bibinfo{author}{Liebel, G.},
  \bibinfo{author}{Raschke, A.}, \bibinfo{author}{Stieglbauer, G.} \&
  \bibinfo{author}{Tichy, M.}
\newblock \bibinfo{editor}{Burgueño, L.} \emph{et~al.} (eds)
  \emph{\bibinfo{title}{Challenges and research directions for successfully
  applying mbe tools in practice}}.
\newblock (eds \bibinfo{editor}{Burgueño, L.} \emph{et~al.})
  \emph{\bibinfo{booktitle}{Proceedings of MODELS 2017 Satellite Event:
  Workshops (ModComp, ME, EXE, COMMitMDE, MRT, MULTI, GEMOC, MoDeVVa, MDETools,
  FlexMDE, MDEbug), Posters, Doctoral Symposium, Educator Symposium, ACM
  Student Research Competition, and Tools and Demonstrations co-located with
  ACM/IEEE 20th International Conference on Model Driven Engineering Languages
  and Systems (MODELS 2017), Austin, TX, USA, September, 17, 2017}}, Vol.
  \bibinfo{volume}{2019} of \emph{\bibinfo{series}{CEUR Workshop Proceedings}},
  \bibinfo{pages}{338--343} (\bibinfo{publisher}{CEUR-WS.org},
  \bibinfo{year}{2017}).
\newblock \urlprefix\url{http://ceur-ws.org/Vol-2019/mdetools_1.pdf}.

\bibitem{david2023collaborative}
\bibinfo{author}{David, I.}, \bibinfo{author}{Aslam, K.},
  \bibinfo{author}{Malavolta, I.} \& \bibinfo{author}{Lago, P.}
\newblock \bibinfo{title}{Collaborative model-driven software engineering—a
  systematic survey of practices and needs in industry}.
\newblock \emph{\bibinfo{journal}{Journal of Systems and Software}}
  \textbf{\bibinfo{volume}{199}}, \bibinfo{pages}{111626}
  (\bibinfo{year}{2023}).

\bibitem{franzago2017collaborative}
\bibinfo{author}{Franzago, M.}, \bibinfo{author}{Di~Ruscio, D.},
  \bibinfo{author}{Malavolta, I.} \& \bibinfo{author}{Muccini, H.}
\newblock \bibinfo{title}{Collaborative model-driven software engineering: a
  classification framework and a research map}.
\newblock \emph{\bibinfo{journal}{IEEE Transactions on Software Engineering}}
  \textbf{\bibinfo{volume}{44}}, \bibinfo{pages}{1146--1175}
  (\bibinfo{year}{2017}).

\bibitem{hidayanto2014impact}
\bibinfo{author}{Hidayanto, A.~N.} \& \bibinfo{author}{Setyady, S.~T.}
\newblock \bibinfo{title}{Impact of collaborative tools utilization on group
  performance in university students.}
\newblock \emph{\bibinfo{journal}{Turkish Online Journal of Educational
  Technology-TOJET}} \textbf{\bibinfo{volume}{13}}, \bibinfo{pages}{88--98}
  (\bibinfo{year}{2014}).

\bibitem{ur2020use}
\bibinfo{author}{UR~RAHMAN, A.}, \bibinfo{author}{Khan, K.},
  \bibinfo{author}{Kamal, S.~W.}, \bibinfo{author}{Naveed, H.} \&
  \bibinfo{author}{Bacha, M.}
\newblock \bibinfo{title}{Use of collaborative tools and modern technologies as
  critical success factor in global software development.}
\newblock \emph{\bibinfo{journal}{Journal on Software Engineering}}
  \textbf{\bibinfo{volume}{15}} (\bibinfo{year}{2020}).

\bibitem{Grischa2014Embedded}
\bibinfo{author}{Liebel, G.}, \bibinfo{author}{Marko, N.},
  \bibinfo{author}{Tichy, M.}, \bibinfo{author}{Leitner, A.} \&
  \bibinfo{author}{Hansson, J.}
\newblock \bibinfo{editor}{Dingel, J.}, \bibinfo{editor}{Schulte, W.},
  \bibinfo{editor}{Ramos, I.}, \bibinfo{editor}{Abrahão, S.} \&
  \bibinfo{editor}{Insfran, E.} (eds) \emph{\bibinfo{title}{Assessing the
  state-of-practice of model-based engineering in the embedded systems
  domain}}.
\newblock (eds \bibinfo{editor}{Dingel, J.}, \bibinfo{editor}{Schulte, W.},
  \bibinfo{editor}{Ramos, I.}, \bibinfo{editor}{Abrahão, S.} \&
  \bibinfo{editor}{Insfran, E.}) , \bibinfo{pages}{166--182}
  (\bibinfo{year}{2014}).

\bibitem{exelmans2022optimistic}
\bibinfo{author}{Exelmans, J.}, \bibinfo{author}{Pietron, J.},
  \bibinfo{author}{Raschke, A.}, \bibinfo{author}{Vangheluwe, H.} \&
  \bibinfo{author}{Tichy, M.}
\newblock \bibinfo{editor}{Dubois, C.} \& \bibinfo{editor}{Cohen, J.} (eds)
  \emph{\bibinfo{title}{Optimistic versioning for conflict-tolerant
  collaborative blended modeling}}.
\newblock (eds \bibinfo{editor}{Dubois, C.} \& \bibinfo{editor}{Cohen, J.})
  \emph{\bibinfo{booktitle}{FPVM 2022: 2nd International Workshop on
  Foundations and Practice of Visual Modeling, July 4--8, 2022, Nantes,
  France}}, Vol. \bibinfo{volume}{3250}, \bibinfo{pages}{1--12}
  (\bibinfo{year}{2022}).

\bibitem{Ryan2000Intrinsic}
\bibinfo{author}{Ryan, R.} \& \bibinfo{author}{Deci, E.}
\newblock \bibinfo{title}{Intrinsic and extrinsic motivations: Classic
  definition and new directions}.
\newblock \emph{\bibinfo{journal}{Contemporary Educational Psychology}}
  \textbf{\bibinfo{volume}{25}}, \bibinfo{pages}{54--67}
  (\bibinfo{year}{2000}).

\bibitem{Kuusinen2016Intrinsic}
\bibinfo{author}{Kuusinen, K.}, \bibinfo{author}{Petrie, H.},
  \bibinfo{author}{Fagerholm, F.} \& \bibinfo{author}{Mikkonen, T.}
\newblock \bibinfo{editor}{Sharp, H.} \& \bibinfo{editor}{Hall, T.} (eds)
  \emph{\bibinfo{title}{Flow, intrinsic motivation, and developer experience in
  software engineering}}.
\newblock (eds \bibinfo{editor}{Sharp, H.} \& \bibinfo{editor}{Hall, T.}) ,
  \bibinfo{pages}{104--117} (\bibinfo{year}{2016}).

\bibitem{Akdur2018Survey}
\bibinfo{author}{Akdur, D.}, \bibinfo{author}{Garousi, V.} \&
  \bibinfo{author}{Demirors, O.}
\newblock \bibinfo{title}{A survey on modeling and model-driven engineering
  practices in the embedded software industry}.
\newblock \emph{\bibinfo{journal}{Journal of Systems Architecture}}
  \textbf{\bibinfo{volume}{91}} (\bibinfo{year}{2018}).

\bibitem{Hutchinson2011Empirical}
\bibinfo{author}{Hutchinson, J.}, \bibinfo{author}{Whittle, J.},
  \bibinfo{author}{Rouncefield, M.} \& \bibinfo{author}{Kristoffersen, S.}
\newblock \bibinfo{editor}{Taylor, R.~N.}, \bibinfo{editor}{Gall, H.} \&
  \bibinfo{editor}{Medvidovic, N.} (eds) \emph{\bibinfo{title}{Empirical
  assessment of mde in industry}}.
\newblock (eds \bibinfo{editor}{Taylor, R.~N.}, \bibinfo{editor}{Gall, H.} \&
  \bibinfo{editor}{Medvidovic, N.}) \emph{\bibinfo{booktitle}{Proceedings of
  the 33rd International Conference on Software Engineering}}, ICSE '11,
  \bibinfo{pages}{471–480} (\bibinfo{publisher}{Association for Computing
  Machinery}, \bibinfo{address}{New York, NY, USA}, \bibinfo{year}{2011}).

\bibitem{Vogelsang2018Embedded}
\bibinfo{author}{Vogelsang, A.}, \bibinfo{author}{Amorim, T.},
  \bibinfo{author}{Pudlitz, F.}, \bibinfo{author}{Gersing, P.} \&
  \bibinfo{author}{Philipps, J.}
\newblock \bibinfo{editor}{Felderer, M.} \emph{et~al.} (eds)
  \emph{\bibinfo{title}{Should i stay or should i go? on forces that drive and
  prevent mbse adoption in the embedded systems industry}}.
\newblock (eds \bibinfo{editor}{Felderer, M.} \emph{et~al.})
  \emph{\bibinfo{booktitle}{Product-Focused Software Process Improvement}},
  \bibinfo{pages}{182--198} (\bibinfo{publisher}{Springer International
  Publishing}, \bibinfo{address}{Cham}, \bibinfo{year}{2017}).

\bibitem{kalantari2023unveiling}
\bibinfo{author}{Kalantari, R.} \& \bibinfo{author}{Lethbridge, T.}
\newblock \bibinfo{editor}{Blouin, A.}, \bibinfo{editor}{Abrahão, S.},
  \bibinfo{editor}{Palanque, P.} \& \bibinfo{editor}{Selic, B.} (eds)
  \emph{\bibinfo{title}{Unveiling developers’ mindset barriers to software
  modeling adoption}}.
\newblock (eds \bibinfo{editor}{Blouin, A.}, \bibinfo{editor}{Abrahão, S.},
  \bibinfo{editor}{Palanque, P.} \& \bibinfo{editor}{Selic, B.})
  \emph{\bibinfo{booktitle}{ACM/IEEE International Conference on Model Driven
  Engineering Languages and Systems Companion}} (\bibinfo{publisher}{IEEE},
  \bibinfo{year}{2023}).
\newblock \bibinfo{note}{Human Factors in Modeling workshop (HuFaMo)}.

\bibitem{petre2013umlInPractice}
\bibinfo{author}{Petre, M.}
\newblock \bibinfo{editor}{Cheng, B. H.~C.} \& \bibinfo{editor}{Pohl, K.} (eds)
  \emph{\bibinfo{title}{Uml in practice}}.
\newblock (eds \bibinfo{editor}{Cheng, B. H.~C.} \& \bibinfo{editor}{Pohl, K.})
  \emph{\bibinfo{booktitle}{Proceedings of the 2013 International Conference on
  Software Engineering}}, ICSE '13, \bibinfo{pages}{722–731}
  (\bibinfo{publisher}{IEEE}, \bibinfo{year}{2013}).

\bibitem{mussbacher2014relevance}
\bibinfo{author}{Mussbacher, G.} \emph{et~al.}
\newblock \bibinfo{editor}{Dingel, J.}, \bibinfo{editor}{Schulte, W.},
  \bibinfo{editor}{Ramos, I.}, \bibinfo{editor}{Abrah{\~a}o, S.} \&
  \bibinfo{editor}{Insfran, E.} (eds) \emph{\bibinfo{title}{The relevance of
  model-driven engineering thirty years from now}}.
\newblock (eds \bibinfo{editor}{Dingel, J.}, \bibinfo{editor}{Schulte, W.},
  \bibinfo{editor}{Ramos, I.}, \bibinfo{editor}{Abrah{\~a}o, S.} \&
  \bibinfo{editor}{Insfran, E.}) \emph{\bibinfo{booktitle}{Model-Driven
  Engineering Languages and Systems}}, \bibinfo{pages}{183--200}
  (\bibinfo{publisher}{Springer International Publishing},
  \bibinfo{year}{2014}).

\bibitem{hall1976beyond}
\bibinfo{author}{Hall, E.~T.}
\newblock \emph{\bibinfo{title}{Beyond culture}}  (\bibinfo{publisher}{Anchor},
  \bibinfo{year}{1976}).

\bibitem{medel2017eliminating}
\bibinfo{author}{Medel, P.} \& \bibinfo{author}{Pournaghshband, V.}
\newblock \bibinfo{editor}{Caspersen, M.~E.}, \bibinfo{editor}{Edwards, S.~H.},
  \bibinfo{editor}{Barnes, T.} \& \bibinfo{editor}{Garcia, D.~D.} (eds)
  \emph{\bibinfo{title}{Eliminating gender bias in computer science education
  materials}}.
\newblock (eds \bibinfo{editor}{Caspersen, M.~E.}, \bibinfo{editor}{Edwards,
  S.~H.}, \bibinfo{editor}{Barnes, T.} \& \bibinfo{editor}{Garcia, D.~D.})
  \emph{\bibinfo{booktitle}{Proceedings of the 2017 ACM SIGCSE technical
  symposium on computer science education}}, \bibinfo{pages}{411--416}
  (\bibinfo{year}{2017}).

\bibitem{baltes2018towards}
\bibinfo{author}{Baltes, S.} \& \bibinfo{author}{Diehl, S.}
\newblock \bibinfo{editor}{Leavens, G.~T.}, \bibinfo{editor}{Garcia, A.} \&
  \bibinfo{editor}{Păsăreanu, C.~S.} (eds) \emph{\bibinfo{title}{Towards a
  theory of software development expertise}}.
\newblock (eds \bibinfo{editor}{Leavens, G.~T.}, \bibinfo{editor}{Garcia, A.}
  \& \bibinfo{editor}{Păsăreanu, C.~S.})
  \emph{\bibinfo{booktitle}{Proceedings of the 2018 26th acm joint meeting on
  european software engineering conference and symposium on the foundations of
  software engineering}}, \bibinfo{pages}{187--200} (\bibinfo{year}{2018}).

\bibitem{catolino2019gender}
\bibinfo{author}{Catolino, G.}, \bibinfo{author}{Palomba, F.},
  \bibinfo{author}{Tamburri, D.~A.}, \bibinfo{author}{Serebrenik, A.} \&
  \bibinfo{author}{Ferrucci, F.}
\newblock \bibinfo{editor}{Kazman, R.} \& \bibinfo{editor}{Pasquale, L.} (eds)
  \emph{\bibinfo{title}{Gender diversity and women in software teams: How do
  they affect community smells?}}
\newblock (eds \bibinfo{editor}{Kazman, R.} \& \bibinfo{editor}{Pasquale, L.})
  \emph{\bibinfo{booktitle}{2019 IEEE/ACM 41st International Conference on
  Software Engineering: Software Engineering in Society (ICSE-SEIS)}},
  \bibinfo{pages}{11--20} (\bibinfo{organization}{IEEE}, \bibinfo{year}{2019}).

\bibitem{NUNES2023102108}
\bibinfo{author}{Nunes, I.}, \bibinfo{author}{Moreira, A.} \&
  \bibinfo{author}{Araujo, J.}
\newblock \bibinfo{title}{Gire: Gender-inclusive requirements engineering}.
\newblock \emph{\bibinfo{journal}{Data \& Knowledge Engineering}}
  \textbf{\bibinfo{volume}{143}}, \bibinfo{pages}{102108}
  (\bibinfo{year}{2023}).

\bibitem{vrieler2021computer}
\bibinfo{author}{Vrieler, T.}, \bibinfo{author}{Nyl{\'e}n, A.} \&
  \bibinfo{author}{Cajander, {\AA}.}
\newblock \bibinfo{title}{Computer science club for girls and boys--a survey
  study on gender differences}.
\newblock \emph{\bibinfo{journal}{Computer Science Education}}
  \textbf{\bibinfo{volume}{31}}, \bibinfo{pages}{431--461}
  (\bibinfo{year}{2021}).

\bibitem{baron2005autism}
\bibinfo{author}{Baron-Cohen, S.} \& \bibinfo{author}{Belmonte, M.~K.}
\newblock \bibinfo{title}{Autism: a window onto the development of the social
  and the analytic brain}.
\newblock \emph{\bibinfo{journal}{Annu. Rev. Neurosci.}}
  \textbf{\bibinfo{volume}{28}}, \bibinfo{pages}{109--126}
  (\bibinfo{year}{2005}).

\bibitem{david2023blended}
\bibinfo{author}{David, I.} \emph{et~al.}
\newblock \bibinfo{title}{Blended modeling in commercial and open-source
  model-driven software engineering tools: A systematic study}.
\newblock \emph{\bibinfo{journal}{Software and Systems Modeling}}
  \textbf{\bibinfo{volume}{22}}, \bibinfo{pages}{415--447}
  (\bibinfo{year}{2023}).

\bibitem{schwartz2012overview}
\bibinfo{author}{Schwartz, S.~H.}
\newblock \bibinfo{title}{An overview of the schwartz theory of basic values}.
\newblock \emph{\bibinfo{journal}{Online readings in Psychology and Culture}}
  \textbf{\bibinfo{volume}{2}}, \bibinfo{pages}{11} (\bibinfo{year}{2012}).

\bibitem{stikkolorum2022studies}
\bibinfo{author}{Stikkolorum, D.~R.}
\newblock \emph{\bibinfo{title}{Studies into interactive didactic approaches
  for learning software design using UML}}.
\newblock Ph.D. thesis, \bibinfo{school}{Leiden University}
  (\bibinfo{year}{2022}).

\bibitem{easterbrook2015modelling}
\bibinfo{author}{Easterbrook, S.}
\newblock \bibinfo{editor}{Lethbridge, T.}, \bibinfo{editor}{Cabot, J.} \&
  \bibinfo{editor}{Egyed, A.} (eds) \emph{\bibinfo{title}{Modelling the climate
  system: Is model-based science like model-based engineering?(keynote)}}.
\newblock (eds \bibinfo{editor}{Lethbridge, T.}, \bibinfo{editor}{Cabot, J.} \&
  \bibinfo{editor}{Egyed, A.}) \emph{\bibinfo{booktitle}{2015 ACM/IEEE 18th
  International Conference on Model Driven Engineering Languages and Systems
  (MODELS)}}, \bibinfo{pages}{1--1} (\bibinfo{organization}{IEEE},
  \bibinfo{year}{2015}).

\bibitem{blair2016grand}
\bibinfo{author}{Blair, G.}
\newblock \bibinfo{editor}{Baudry, B.}, \bibinfo{editor}{Combemale, B.},
  \bibinfo{editor}{Kienzle, J.} \& \bibinfo{editor}{Pretschner, A.} (eds)
  \emph{\bibinfo{title}{Grand challenges, grand responses?(keynote)}}.
\newblock (eds \bibinfo{editor}{Baudry, B.}, \bibinfo{editor}{Combemale, B.},
  \bibinfo{editor}{Kienzle, J.} \& \bibinfo{editor}{Pretschner, A.})
  \emph{\bibinfo{booktitle}{2016 ACM/IEEE 19th International Conference on
  Model Driven Engineering Languages and Systems (MODELS)}}
  (\bibinfo{organization}{ACM}, \bibinfo{year}{2016}).

\end{thebibliography}

\end{document}